\documentclass[a4paper]{llncs}
\usepackage{subcaption}

\usepackage{paralist,url}
\usepackage{graphicx}
\usepackage{pdfpages}
\usepackage[title]{appendix}
\usepackage{caption}
\usepackage{subcaption}
\usepackage{enumitem}
\usepackage{float}
\floatstyle{ruled}
\newfloat{program}{t}{lop}
\floatname{program}{Program}

\let \citep = \cite
\newcommand{\Comment}[1]{}

\title{A Proof Strategy Language and Proof Script Generation for Isabelle/HOL}
\author{Yutaka Nagashima \and Ramana Kumar}
\institute{Data61, CSIRO / UNSW}

\newcommand{\lang}{\texttt{PSL}}

\newcommand{\strategy}{\texttt{strategy}}
\newcommand{\findproof}{\texttt{find\_proof}}
\newcommand{\tryhard}{\texttt{try\_hard}}

\newcommand{\THEN}{\texttt{THEN}}
\newcommand{\APPEND}{\texttt{APPEND}}
\newcommand{\ORELSE}{\texttt{ORELSE}}
\newcommand{\REPEAT}{\texttt{REPEAT}}
\newcommand{\Thens}{\texttt{Thens}}
\newcommand{\Alts}{\texttt{Alts}}
\newcommand{\Ors}{\texttt{Ors}}
\newcommand{\Alt}{\texttt{Alt}}
\newcommand{\Or}{\texttt{Or}}
\newcommand{\POrs}{\texttt{POrs}}
\newcommand{\PAlts}{\texttt{PAlts}}
\newcommand{\Repeat}{\texttt{Repeat}}

\newcommand{\etal}{\textit{et\,al.}}

\newcommand{\lemmTwo}{\texttt{safe\_trans}}
\newcommand{\pgOne}{\texttt{1:"ps\_safe \textit{p s}"}}
\newcommand{\pgTwo}{\texttt{2:"valid\_tran \textit{p s s' c}"}}
\newcommand{\pgThree}{\texttt{3:"ps\_safe \textit{p s'}"}}
\newcommand{\strOne}{\texttt{DInductAuto}}

\newcommand{\induct}{\texttt{induct}}
\newcommand{\auto}{\texttt{auto}}
\newcommand{\DInduct}{\texttt{Dynamic(Induct)}}
\newcommand{\Auto}{\texttt{Auto}}
\newcommand{\issolved}{\texttt{is\_solved}}
\newcommand{\fastforce}{\texttt{fastforce}}

\newcommand{\IsSolved}{\texttt{IsSolved}}
\newcommand{\Defer}{\texttt{Defer}}
\newcommand{\sledgehammer}{\texttt{sledgehammer}}
\newcommand{\quickcheck}{\texttt{quickcheck}}
\newcommand{\nitpick}{\texttt{nitpick}}

\newcommand{\succeed}{\texttt{succeed}}
\newcommand{\fail}{\texttt{fail}}
\newcommand{\Skip}{\texttt{Skip}}
\newcommand{\Fail}{\texttt{Fail}}
\newcommand{\mzero}{\texttt{mzero}}

\newcommand{\transfer}{\texttt{transfer}}

\newcommand{\Eisbach}{\textit{Eisbach}}
\newcommand{\IsaPlanner}{\textit{IsaPlanner}}

\newcommand{\grind}{\texttt{grind}}
\newcommand{\PVS}{\textit{PVS}}
\newcommand{\SEPIA}{\textit{SEPIA}}
\newcommand{\ACL}{\textit{ACL2}}

\newcommand{\Isar}{\textit{Isar}}
\newcommand{\g}{\textit{g}}
\newcommand{\xs}{\textit{xs}}
\newcommand{\ys}{\textit{ys}}
\newcommand{\zs}{\textit{zs}}
\newcommand{\dfsapp}{\texttt{dfs\_app}}
\newcommand{\someinduct}{\texttt{DInductAuto}}
\newcommand{\moreinduct}{\texttt{some\_induct2}}

\newcommand{\checkpickdef}
  {\texttt{Thens} \texttt{[Quickcheck,} \texttt{Defer]}}
\newcommand{\eval}{\texttt{eval}}
\newcommand{\two}{\texttt{2}}
\newcommand{\comment}[1]{}
\newcommand{\PThenOne}{\texttt{PThenOne}}
\newcommand{\PThenAll}{\texttt{PThenAll}}
\newcommand{\User}{\texttt{User}}

\begin{document}

  \maketitle

  \begin{abstract}
   We introduce a language, PSL, designed to capture high level proof strategies in Isabelle/HOL.
   Given a strategy and a proof obligation,
   PSL's runtime system generates and combines various tactics
   to explore a large search space with low memory usage.
   Upon success, PSL generates an efficient proof script, which bypasses a large part of the proof search.
   We also present PSL's monadic interpreter to show that
   the underlying idea of PSL is transferable to other ITPs.
  \end{abstract}

\section{Introduction}\label{s:intro}

\begin{figure}[b]
    \centering
    \begin{subfigure}[b]{0.48\textwidth}
        \includegraphics[width=\textwidth]{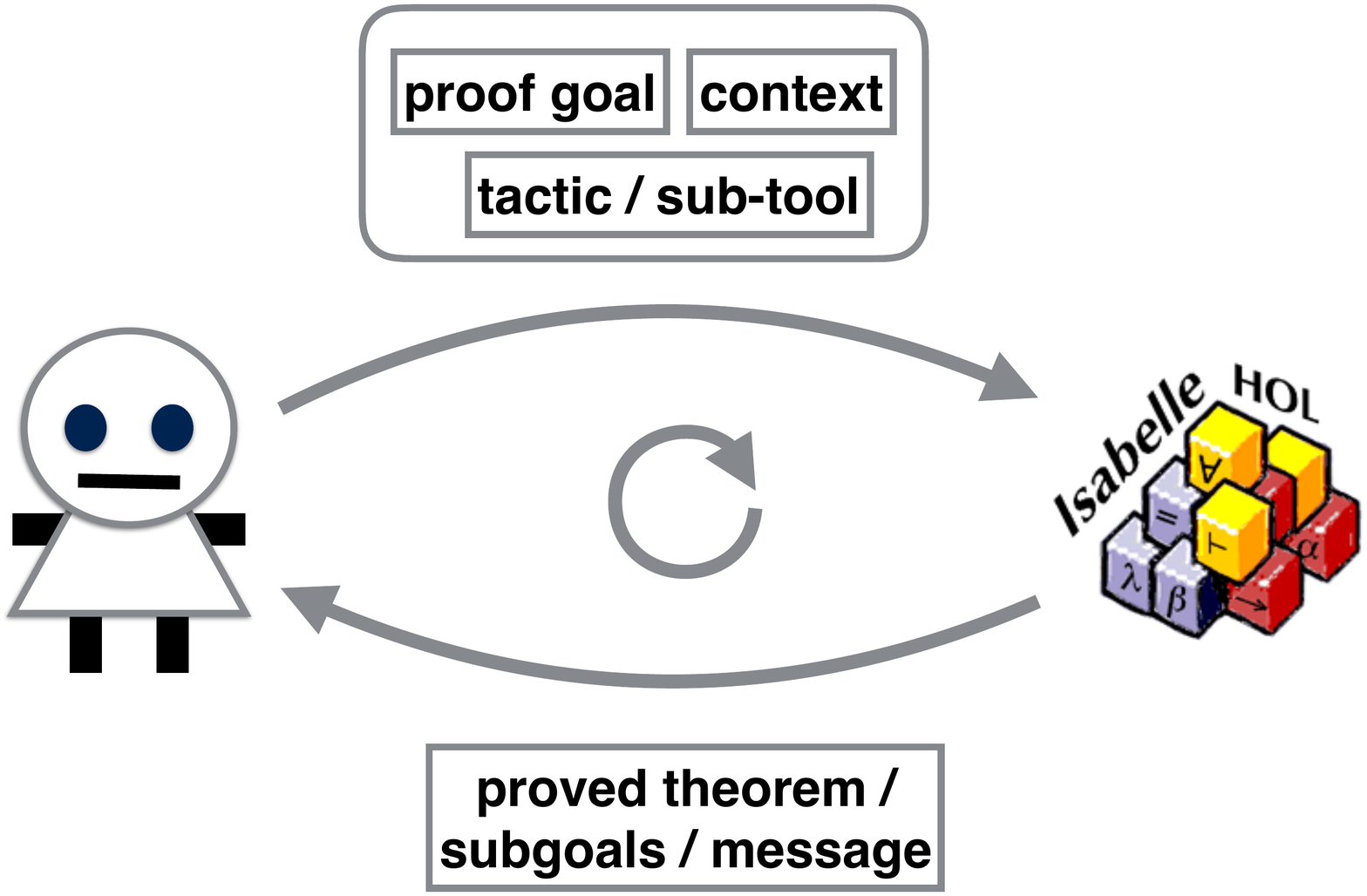}
        \caption{Standard proof attempt}
        \label{fig:wo_psl}
    \end{subfigure}
    ~ 
    \begin{subfigure}[b]{0.48\textwidth}
        \includegraphics[width=\textwidth]{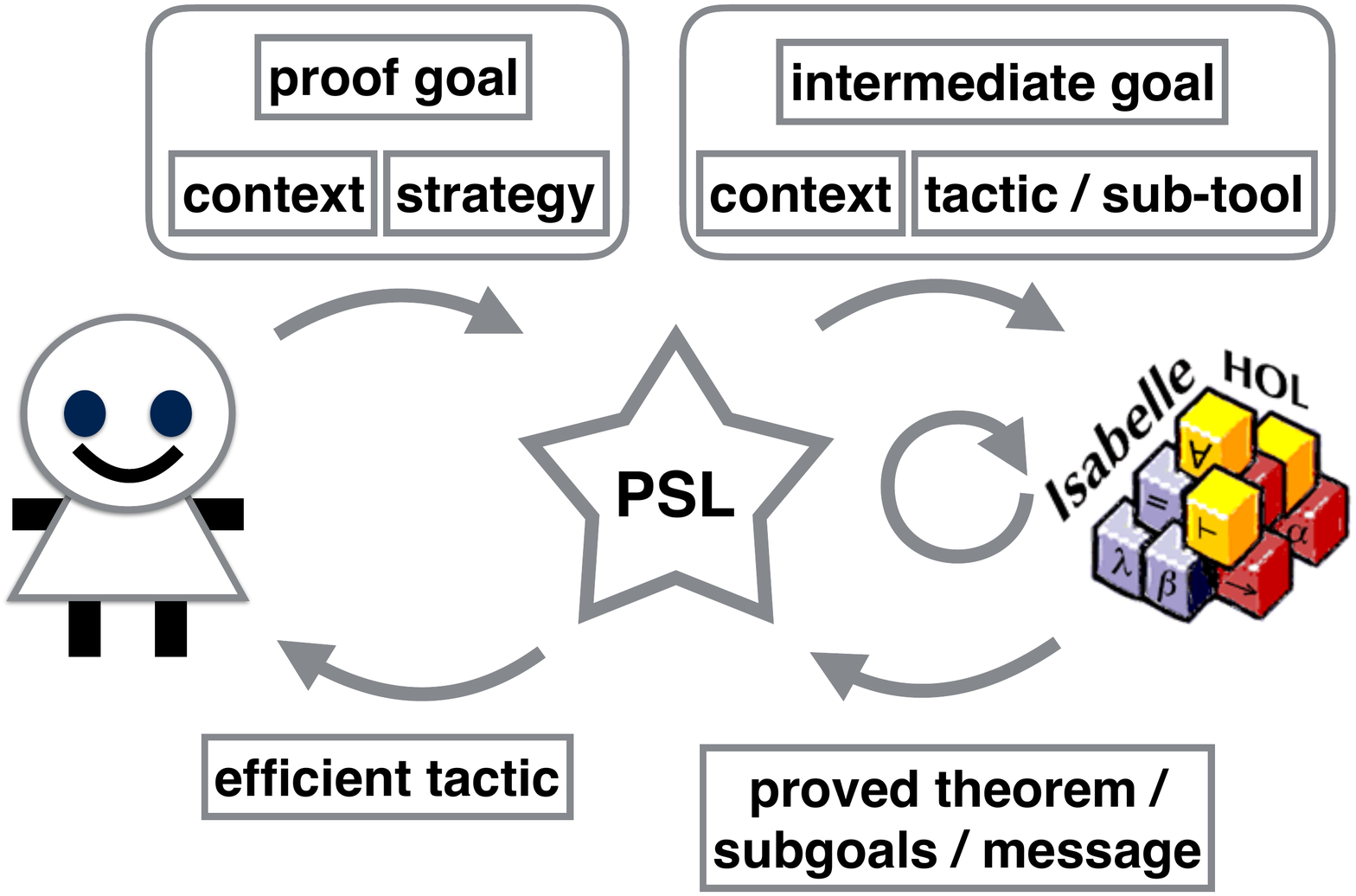}
        \caption{Proof attempt with \texttt{PSL}}
        \label{fig:with_psl}
    \end{subfigure}
    \caption{Comparison of proof development processes}\label{fig:process}
\end{figure}

Currently, users of interactive theorem provers (ITPs) spend too much time iteratively interacting with their ITP
to manually specialise and combine tactics as depicted in Fig. \ref{fig:wo_psl}.
This time consuming process requires expertise in the ITP,
making ITPs more esoteric than they should be.
The integration of powerful automated theorem provers (ATPs) into ITPs ameliorates this problem significantly;
however, the exclusive reliance on general purpose ATPs makes it hard to exploit
users' domain specific knowledge,
leading to combinatorial explosion even for conceptually straight-forward conjectures.

To address this problem,
we introduce \lang{}, a programmable, extensible, meta-tool based framework,
to Isabelle/HOL \citep{DBLP:books/sp/NipkowPW02}.
We provide \lang{} (available on GitHub \cite{PSL}) as a language,
so that its users can encode \textit{proof strategies},
abstract descriptions of how to attack proof obligations,
based on their intuitions about a conjecture.
When applied to a proof obligation,
\lang{}'s runtime system creates and combines several tactics
based on the given proof strategy.
This makes it possible to explore a larger search space
than has previously been possible with conventional tactic languages,
while utilising users' intuitions on the conjecture.

We developed \lang{} to utilise engineers' downtime:
with \lang{}, we can run an automatic proof search for hours
while we are attending meetings, sleeping, or reviewing papers.
\lang{} makes such expensive proof search possible
on machines with limited memory:
\lang{}'s runtime truncates failed proof attempts as soon as it backtracks
to minimise its memory usage.

Furthermore,
\lang{}'s runtime system attempts to generate efficient proof scripts
from a given strategy
by searching for the appropriate specialisation and combination of tactics
for a particular conjecture
without direct user interaction, as illustrated in Fig. \ref{fig:with_psl}.
Thus, \lang{} not only reduces the initial labour cost of theorem proving,
but also keeps proof scripts interactive and maintainable
by reducing the execution time of subsequent proof checking.

In Isabelle, \sledgehammer{} adopts a similar approach
\citep{DBLP:journals/jfrea/BlanchetteKPU16}.
It exports a proof goal to various external ATPs
and waits for them to find a proof.
If the external provers find a proof,
\sledgehammer{} tries to reconstruct an efficient proof script in Isabelle
using hints from the ATPs.
\sledgehammer{} is often more capable than most tactics
but suffers from discrepancies between the different provers and logics used.
While we integrated \sledgehammer{} as a sub-tool in \lang{},
\lang{} conducts a search using Isabelle tactics,
thus avoiding the problems arising from the discrepancies and proof reconstruction.

The underlying implementation idea in \lang{} is the monadic interpretation of proof strategies, which we introduce in Section \ref{s:psm}.
We expect this prover-agnostic formalization
brings the following strengths of \lang{}
to other ITPs such as Lean \cite{DBLP:conf/cade/MouraKADR15}
and Coq \cite{09thecoq}:
\begin{itemize}[noitemsep,nolistsep]
\item runtime tactic generation based on user-defined procedures,
\item memory-efficient large-scale proof search, and
\item efficient-proof-script generation for proof maintenance.
\end{itemize}

\section{Background}\label{background}
Interactive theorem proving can be seen as the exploration of a search tree.
Nodes of the tree represent proof states.
Edges represent applications of tactics, which transform the proof state.
The behaviour of tactics are, in general, not completely predictable
since they are context sensitive:
they behave differently depending on the information stored in background proof contexts.
A proof context contains the information such as
constants defined so far and auxiliary lemmas proved prior to the step.
Therefore, the shape of the search tree is not known in advance.

\begin{figure}
    \centering
    \begin{subfigure}[b]{0.48\textwidth}
        \includegraphics[width=\textwidth]{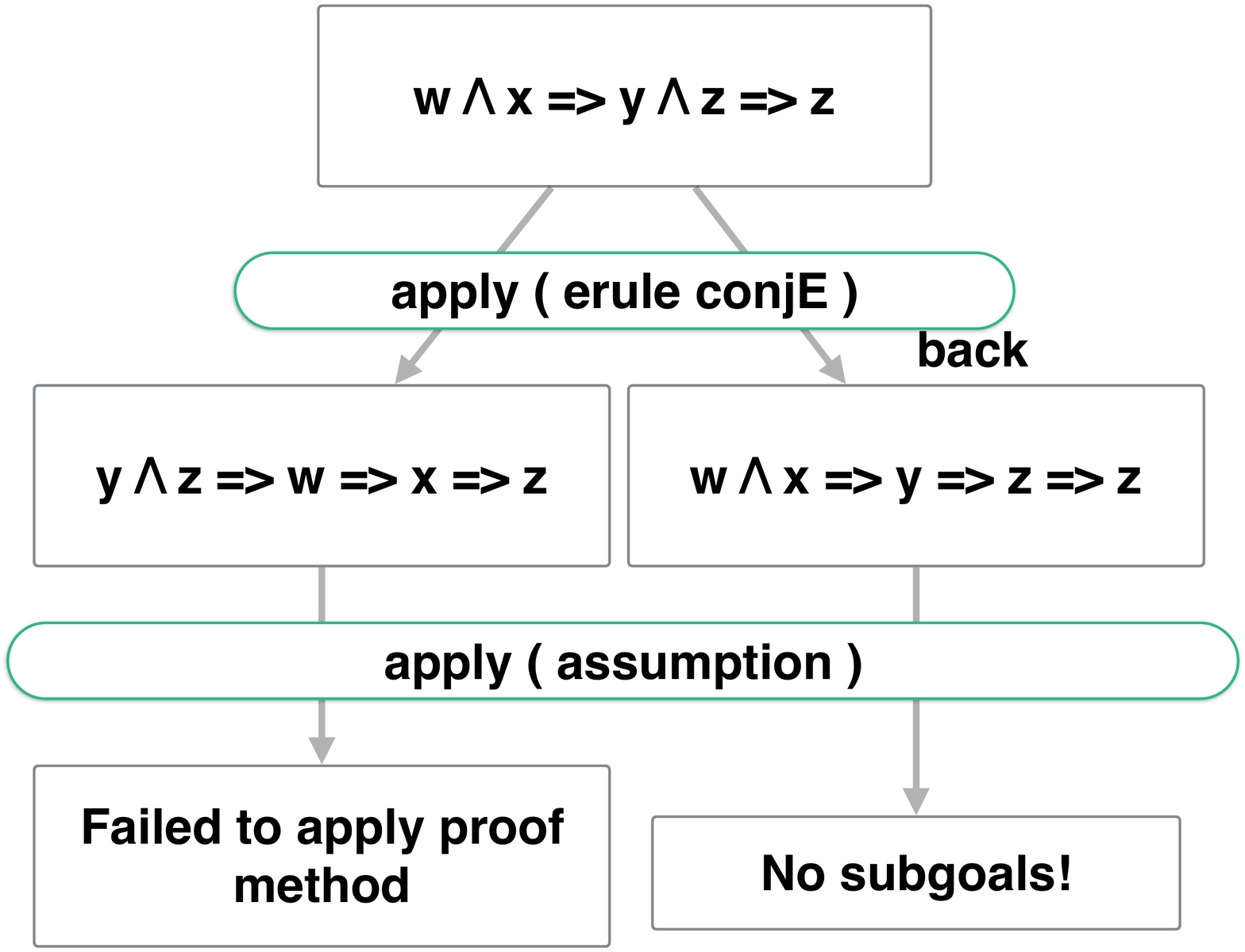}
        \caption{External view}
        \label{fig:external}
    \end{subfigure}
    ~ 
    \begin{subfigure}[b]{0.48\textwidth}
        \includegraphics[width=\textwidth]{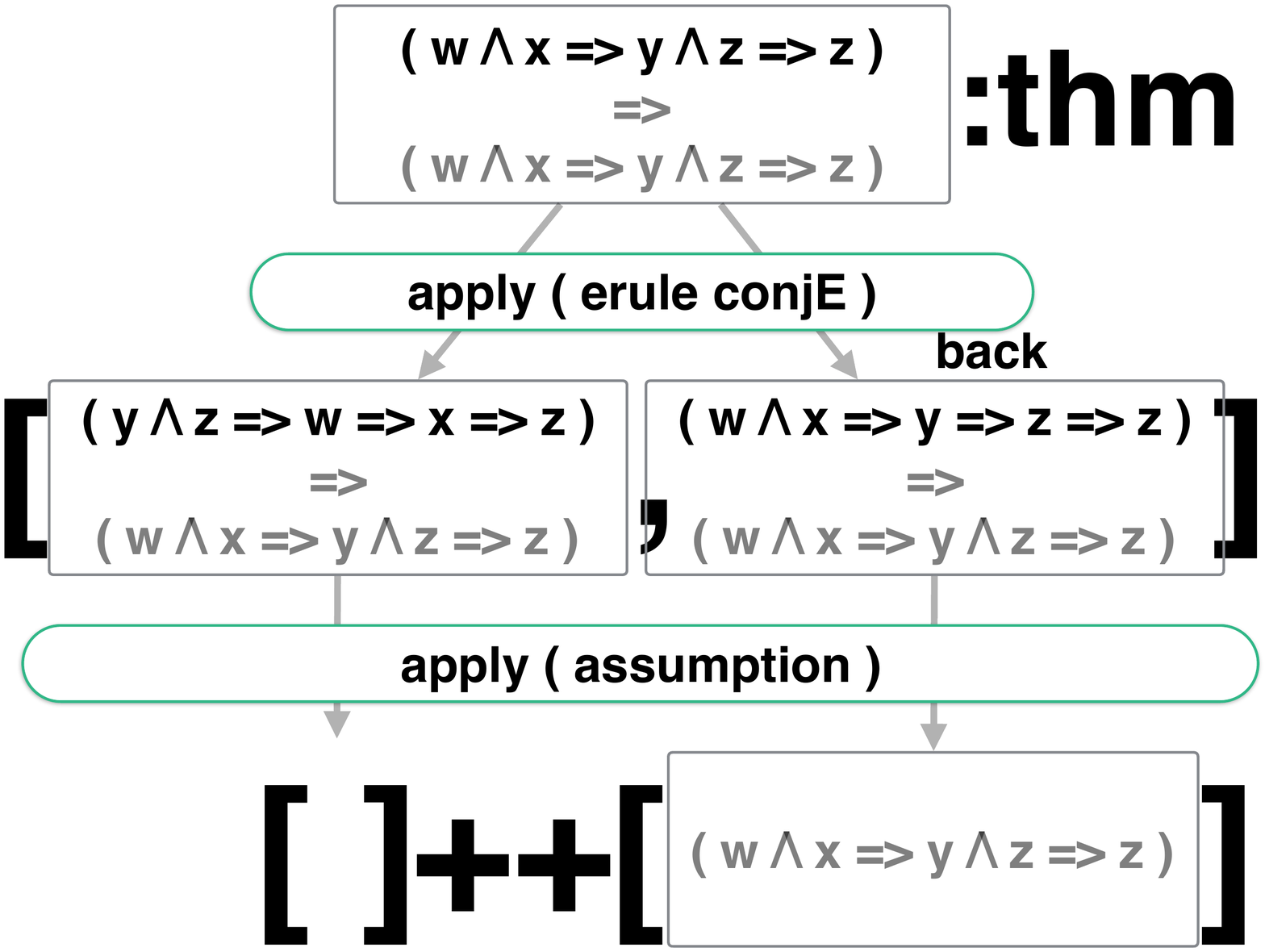}
        \caption{Internal view}
        \label{fig:internal}
    \end{subfigure}
    \caption{External and internal view of proof search tree.}
    \label{fig:example_search_tree}
\end{figure}

The goal is to find a node representing a solved state: one in which the proof is complete.
The search tree may be infinitely wide and deep,
because there are endless variations of tactics that may be tried at any point.
The goal for a \lang{} strategy is
to direct an automated search of this tree to find a solved state;
\lang{} will reconstruct an efficient path to this state as a human-readable proof script.

Fig. \ref{fig:external} shows an example of proof search.
At the top, the tactic \verb|erule conjE| is applied to the proof obligation
\texttt{w$\wedge$x $\Rightarrow$ y$\wedge$z $\Rightarrow$ z}.
This tactic invocation produces two results,
as there are two places to apply conjunction elimination.
Applying conjunction elimination to \texttt{w$\wedge$x} returns
the first result,
while doing so to \texttt{y$\wedge$z} produces the second result.
Subsequent application of proof by \verb|assumption| can discharge the second result;
however, \verb|assumption| does not discharge the first one
since the \verb|z| in the assumptions is still hidden by the conjunction.
Isabelle's proof language, \Isar{}, returns the first result by default,
but users can access the subsequent results using the keyword \verb|back|.

Isabelle represents this non-deterministic behaviour of tactics using lazy sequences:
tactics are functions of type \verb|thm -> [thm]|,
where \texttt{[$\cdot$]} denotes a (possibly infinite) lazy sequence
\cite{DBLP:journals/corr/cs-LO-9301105}.
Fig. \ref{fig:internal} illustrates
how Isabelle internally handles the above example.
Each proof state is expressed as a (possibly nested) implication
which assumes proof obligations to conclude the conjecture.
One may complete a proof by removing these assumptions using tactics.
Tactic failure is represented as an empty sequence,
which enables backtracking search
by combining multiple tactics in a row \cite{DBLP:conf/fpca/Wadler85}.
For example, one can write
\verb|apply|\verb|(erule|  \verb|conjE,| \verb|assumption)|
using the sequential combinator \texttt{,} (comma) in \Isar{};
this tactic traverses the tree using backtracking
and discharges the proof obligation
without relying on the keyword \verb|back|.

The search tree grows wider when choosing between multiple tactics,
and it grows deeper when tactics are combined sequentially.
In the implementation language level,
the tactic combinators in Isabelle include
\texttt{THEN} for sequential composition
(corresponding to \verb|,| in \Isar{}),
\texttt{APPEND} for non-deterministic choice,
\texttt{ORELSE} for deterministic choice, and
\texttt{REPEAT} for iteration.

Isabelle/HOL comes with several default tactics such as
\verb|auto|, \verb|simp|, \verb|induct|, \verb|rule|, and \verb|erule|.
When using tactics, proof authors often have to adjust tactics
using \textit{modifier}s for each proof obligation.
\succeed{} and \fail{} are special tactics:
\succeed{} takes a value of type \texttt{thm}, wraps it in a lazy sequence, and returns it
without modifying the value.
\fail{} always returns an empty sequence. 

\section{Syntax of \lang{}}\label{s:syntax}
The following is the syntax of \lang{}.
We made \lang{}'s syntax similar to that of Isabelle's tactic language
aiming at the better usability for users who are familiar with Isabelle's tactic language.

\begin{verbatim}
strategy = default | dynamic | special | subtool | compound
default  = Simp | Clarsimp | Fastforce | Auto | Induct
         | Rule | Erule | Cases | Coinduction | Blast
dynamic  = Dynamic (default)
special  = IsSolved | Defer | IntroClasses | Transfer
         | Normalization | Skip | Fail | User <string>
subtool  = Hammer | Nitpick | Quickcheck
compound = Thens [strategy] | Ors [strategy] | Alts [strategy]
         | Repeat(strategy) | RepeatN (strategy)
         | POrs  [strategy] | PAlts [strategy]
         | PThenOne [strategy] | PThenAll [strategy]
         | Cut int (strategy)
\end{verbatim}

\noindent
The \texttt{default} strategies correspond to Isabelle's default tactics without arguments,
while \texttt{dynamic} strategies correspond to Isabelle's default tactics
that are specialised for each conjecture.
Given a \texttt{dynamic} strategy and conjecture,
the runtime system generates variants of the corresponding Isabelle tactic.
Each of these variants is specialised for the conjecture
with a different combination of promising arguments found in the conjecture and its proof context.
It is the purpose of the \lang{} runtime system to select the right combination automatically.

\texttt{subtool} represents Isabelle tools
such as \sledgehammer{} \citep{DBLP:journals/jfrea/BlanchetteKPU16}
and counterexample finders.
The \texttt{compound} strategies capture the notion of tactic combinators:
\Thens{} corresponds to \THEN,
\Ors{} to \ORELSE{},
\Alts{} to \APPEND{}, and
\Repeat{} to \REPEAT{}.
\POrs{} and \PAlts{} are similar to \Ors{} and \Alts{}, respectively,
but they admit parallel execution of sub-strategies.
\PThenOne{} and \PThenAll{} take exactly two sub-strategies,
combine them sequentially
and apply the second sub-strategy to the results of the first sub-strategy in parallel
in case the first sub-strategy returns multiple results.
Contrary to \PThenAll{}, \PThenOne{} stops its execution as soon as
it produces one result from the second sub-strategy.
Users can integrate user-defined tactics, including those written
in \Eisbach{} \citep{DBLP:conf/itp/MatichukWM14},
into \lang{} strategies using \User{}.
\verb|Cut| limits the degree of non-determinism within a strategy.

In the following, we explain how to write strategies and
how \lang{}'s runtime system interprets strategies with examples.

\section{\lang{} by Example}\label{s:example}
\begin{figure}[t]
  \centering
  \includegraphics[width=0.92\textwidth]{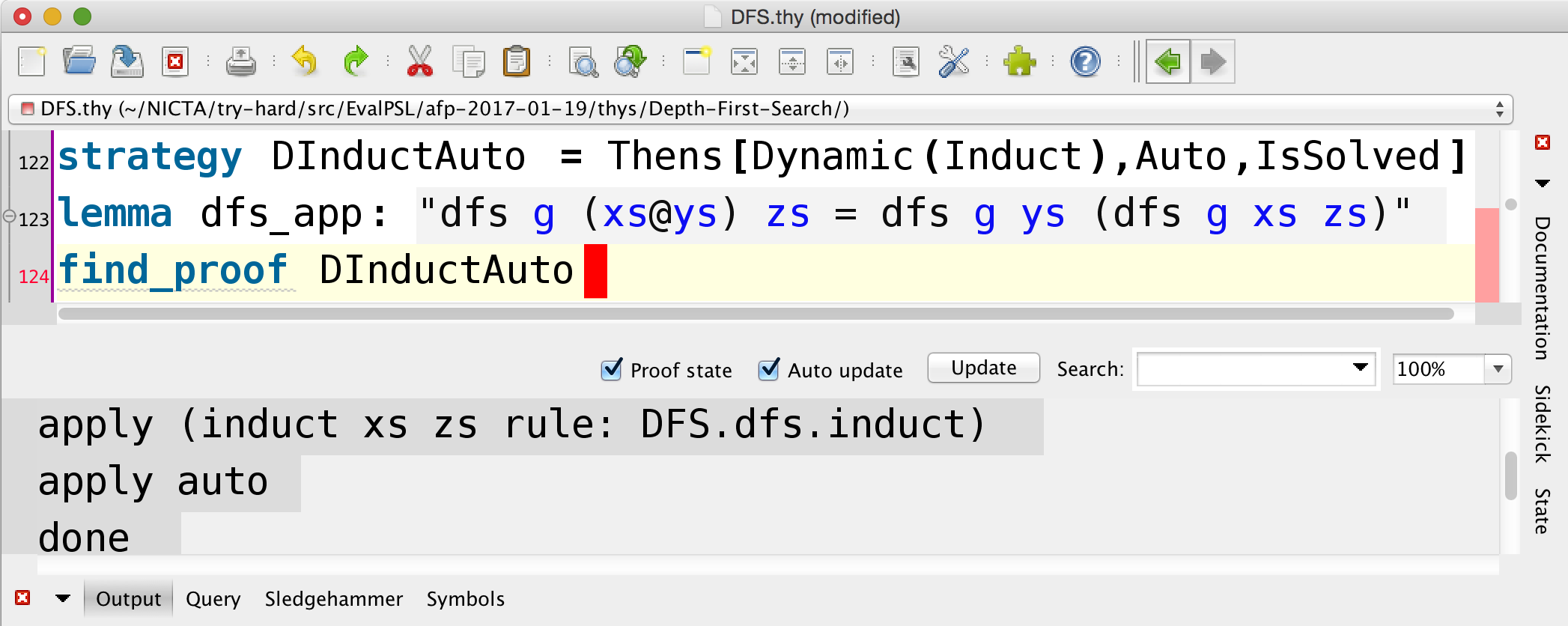}
  \caption{Screenshot for Example 1.} \label{figure:tall}
\end{figure}

\paragraph{Example 1.}
For our first example,
we take the following lemma about depth-first search from an entry
\citep{Depth-First-Search-AFP}.\\
\\
\texttt{lemma dfs\_app: "dfs \g{} (\xs{} @ \ys{}) \zs{} =
dfs \g{} \ys{} (dfs \g{} \xs{} \zs{})"}\\
\\
where \verb|dfs| is a recursively defined function for depth-first search.
As \verb|dfs| is defined recursively,
it is natural to expect that its proof involves some sort of mathematical induction.
However, we do not know exactly how we should conduct mathematical induction here;
therefore, we describe this rough idea as a proof strategy, \someinduct{},
with the keyword \strategy{},
and apply it to \dfsapp{} with the keyword \findproof{} as depicted in Fig. \ref{figure:tall}.
Invoked by \findproof{}, \lang{}'s runtime system interprets \someinduct{}.
For example, it interprets \Auto{} as Isabelle's default tactic, \auto{}.

The interpretation of \DInduct{} is more involved:
the runtime generates tactics using the information in \dfsapp{}
and its background context.
First, \lang{} collects the free variables (noted in \textit{italics} above) in \dfsapp{}
and applicable induction rules stored in the context.
\lang{} uses the set of free variables to specify two things:
on which variables instantiated tactics conduct mathematical induction,
and which variables should be generalised in the induction scheme.
The set of applicable rules are used to specify which rules to use.
Second, \lang{} creates the powerset out of the set of all possible modifiers.
Then, it attempts to instantiate a variant of the induct tactic for each subset of modifiers.
Finally, it combines all the variants of \induct{} with unique results using \APPEND{}.
In this case, \lang{} tries to generate 4160 induct tactics for \dfsapp{}
by passing several combinations of modifiers to Isabelle;
however, Isabelle cannot produce valid induction schemes for some combinations,
and some combinations lead to the same induction scheme.
The runtime removes these,
focusing on the 223 unique results.
\lang{}'s runtime combine these tactics with \auto{} using \THEN{}.

\lang{}'s runtime interprets \verb|IsSolved| as the \issolved{} tactic, which
checks if any proof obligations are left or not.
If obligations are left,
\issolved{} behaves as \fail{}, triggering backtracking.
If not,
\issolved{} behaves as \succeed{},
allowing the runtime to stop the search.
This is how \someinduct{} uses \IsSolved{} to ensure that
no sub-goals are left before returning an efficient proof script.
For \dfsapp{},
\lang{} interprets \someinduct{} as
the following tactic:
\vspace*{1em}\\
\noindent
\texttt{(induct1 APPEND induct2 APPEND\ldots) THEN auto THEN is\_solved}
\vspace*{1em}\\
\noindent
where \verb|induct_n|s are
variants of the induct tactic specialised with modifiers.

\begin{figure}
    \centering
    \begin{subfigure}[b]{0.48\textwidth}
        \includegraphics[width=\textwidth]{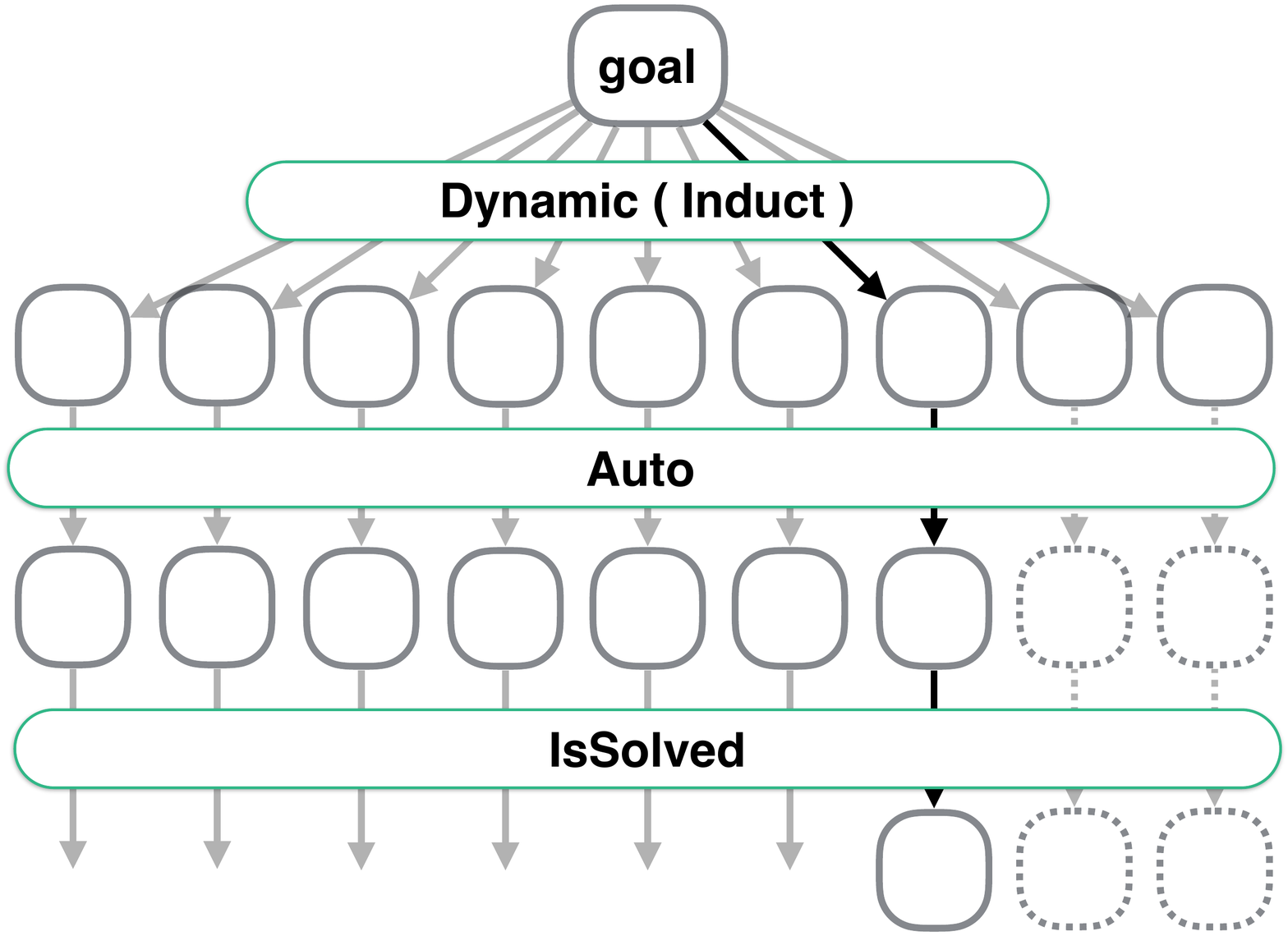}
        \caption{\someinduct}
        \label{fig:someinduct}
    \end{subfigure}
    ~ 
    \begin{subfigure}[b]{0.48\textwidth}
        \includegraphics[width=\textwidth]{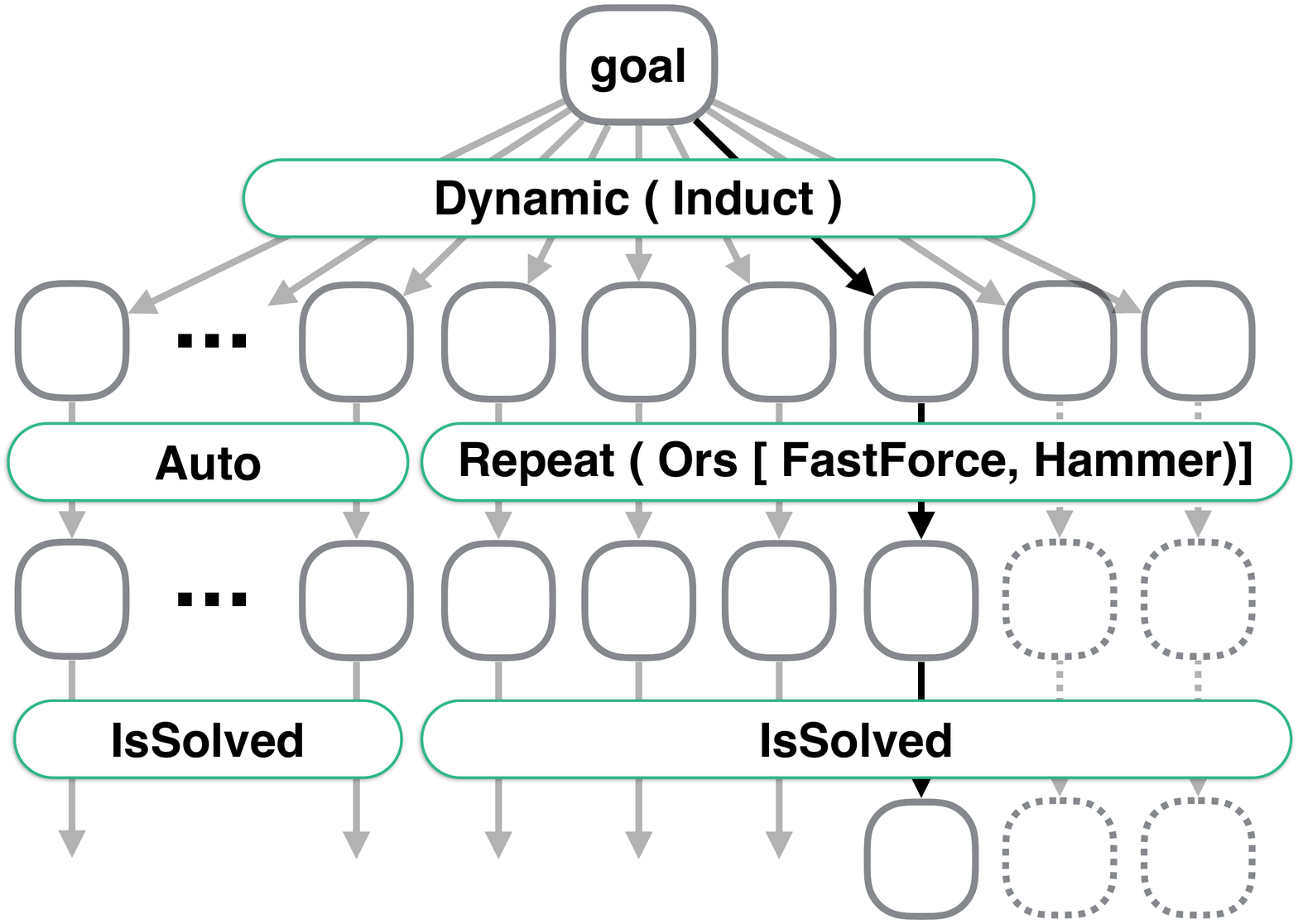}
        \caption{\moreinduct}
        \label{fig:moreinduct}
    \end{subfigure}
    \caption{Proof search tree for \texttt{some\_induct}}
    \label{fig:dynamic_induct}
\end{figure}

Within the runtime system,
Isabelle first applies \texttt{induct1} to \dfsapp{},
then \auto{} to the resultant proof obligations.
Note that each \induct{} tactic and \auto{} is deterministic:
it either fails or returns a lazy sequence with a single element.
However, combined together with \APPEND{},
the numerous variations of \induct{} tactics \textit{en mass} are non-deterministic:
if \issolved{} finds remaining proof obligations,
Isabelle backtracks to the next \induct{} tactic,
\texttt{induct2} and repeats this process
until either it discharges all proof obligations or
runs out of the variations of \induct{} tactics.
The numerous variants of \induct{} tactics from \strOne{} allow
Isabelle to explore a larger search space than its naive alternative,
\texttt{induct THEN auto}, does.
Fig. \ref{fig:someinduct} illustrates this search procedure.
Each edge and curved square represents a tactic application and a proof state, respectively,
and edges leading to no object stand for tactic failures.
The dashed objects represent possible future search space,
which \lang{} avoids traversing by using lazy sequences.

The larger search space specified by \someinduct{} leads to a longer search time.
\lang{} addresses this performance problem by tracing Isabelle's proof search:
it keeps a log of successful proof attempts while removing backtracked proof attempts.
The monadic interpretation discussed in Section \ref{s:psm} let
\lang{} remove failed proof steps as soon as it backtracks.
This minimises \lang{} memory usage, making it applicable to hours of expensive automatic proof search.
Furthermore, since \lang{} follows Isabelle's execution model based on lazy sequences,
it stops proof search as soon as it finds a specialisation and combination of tactics,
with which Isabelle can pass the no-proof-obligation test imposed by \issolved{}.

We still need a longer search time with \lang{}, but only once:
upon success, \lang{} converts the log of successful attempts into an efficient proof script,
which bypasses a large part of proof search.
For \dfsapp{}, \lang{} generates the following proof script from \someinduct{}.
\begin{verbatim}
apply (induct xs zs rule: DFS.dfs.induct) apply auto done
\end{verbatim}
We implemented \lang{} as an Isabelle theory;
to use it, \lang{} users only have to import the relevant theory files to use \lang{}
to their files.
Moreover, we have integrated \lang{} into Isabelle/Isar, Isabelle's proof language,
and Isabelle/jEdit, its standard editor.
This allows users to define and invoke their own proof strategies inside their ongoing proof attempts,
as shown in Figure \ref{figure:tall};
and if the proof search succeeds \lang{} presents a proof script in jEdit's output panel,
which users can copy to the right location with one click.
All generated proof scripts are independent of \lang{},
so users can maintain them without \lang{}.

\paragraph*{Example 2.}
\someinduct{} is able to pick up the right induction scheme
for relatively simple proof obligations using backtracking search.
However, in some cases even if \lang{} picks the right induction scheme,
\auto{} fails to discharge the emerging sub-goals.
In the following, we define \verb|InductHard|,
a more powerful strategy based on mathematical induction,
by combining \verb|Dynamic|\verb|(Induct)| with more involved sub-strategies
to utilise external theorem provers.

\begin{verbatim}
strategy SolveAllG = Thens[Repeat(Ors[Fastforce,Hammer]),IsSolved]
strategy PInductHard = PThenOne[Dynamic(Induct),SolveAllG]
strategy InductHard = Ors[DInductAuto, PInductHard]
\end{verbatim}

\lang{}'s runtime system interprets 
\verb|Fastforce| and \verb|Hammer| as the \fastforce{} tactic and \sledgehammer{}, respectively.
Both \fastforce{} and \sledgehammer{} try to discharge the first sub-goal only
and return an empty sequence if they cannot discharge the sub-goal.
\Comment{This is contrary to \auto{},
which works on all the sub-goals and returns a lazy sequence with one element
(representing possibly multiple sub-goals)
even if it does not discharge any sub-goals completely.
Furthermore, \auto{} is an un-safe tactic:
it can transform the proof obligation into an unprovable form.
Hence, \verb|AutoHammers| wraps \Auto{} within \verb|Subgoal| and \verb|IsSolved|,
forcing \auto{} to focus on the first sub-goal to protect the provability of other sub-goals.
Then, it applies \sledgehammer{} repeatedly to the sub-goals
arising from the application of \auto{}.
If \IsSolved{} in \verb|AutoHammers| succeeds
confirming there is no sub-goal left in this focused area,
it produces the Isar command \verb|done| to conclude the scope of focusing,
bringing back the ignored sub-goals into the scope of \lang{}.}

The repetitive application of \sledgehammer{} would be very time consuming.
We mitigate this problem using \Ors{} and \verb|PThenOne|.
Combined with \Ors{}, \lang{} executes \verb|PInductHard|
only if \verb|DInductAuto| fails.
When \verb|PInductHard| is called,
it first applies \verb|Dynamic(Induct)|, producing various induction schemes and multiple results.
Then, \verb|SolveAllG| tries to discharge these results in parallel.
The runtime stops its execution when \verb|SolveAllG| returns at least one result representing a solved state.
We apply this strategy to the following conjecture, which
states the two versions of depth first search programs (\verb|dfs2| and \verb|dfs|)
return the same results given the same inputs.\\
\noindent
\\
\texttt{lemma "dfs2 \g{} \xs{} \ys{} = dfs \g{} \xs{} \ys{}"}
\\

\noindent
Then, our machine with 28 cores returns the following script within 3 minutes:
\Comment{\begin{verbatim}
apply (induct) apply fastforce subgoal apply auto
using local.sorted.Cons local.sorted_append apply fastforce
done done
\end{verbatim}}
\begin{verbatim}
apply (induct xs ys rule: DFS.dfs2.induct)
apply fastforce apply (simp add: dfs_app) done
\end{verbatim}

Fig. \ref{fig:moreinduct} roughly shows how the runtime system
found this proof script.
The runtime first tried to find a complete proof as in Example 1, but without much success.
Then, it interpreted \verb|PInductHard|.
While doing so,
it found that
induction on \xs{} and \ys{} using \verb|DFS.dfs2.induct| leads to two sub-goals
both of which can be discharged either by \fastforce{} or \sledgehammer{}.
For the second sub-goal,
\sledgehammer{} found out that
the result of \textit{Example 1} can be used as an auxiliary lemma to prove this conjecture.
Then, it returns an efficient-proof-script \verb|(simp add: dfs_app)| to \verb|PSL|,
before \lang{} combines this with other parts and prints the complete proof script.

\paragraph{Example 3.}\label{incomplete_proof}
In the previous examples,
we used \IsSolved{} to get a complete proof script from \lang{}.
In Example 3, we show how to generate incomplete but useful proof scripts,
using \Defer{}.
Incomplete proofs are specially useful
when ITP users face numerous proof obligations,
many of which are within the reach of high-level proof automation tools,
such as \sledgehammer{},
but a few of which are not.

Without \lang{}, Isabelle users had to manually invoke \sledgehammer{} several times
to find out which proof obligations \sledgehammer{} can discharge.
We developed a strategy, \verb|HamCheck|, to automate this time-consuming process.
The following shows its definition and a use case simplified for illustrative purposes.\\
\\
\noindent
\verb|strategy HamCheck = RepeatN(Ors[Hammer,Thens[Quickcheck,Defer]])|\\
\texttt{lemma \lemmTwo{}: shows\\
\hspace*{0em}\pgOne{} and \pgTwo{} and \pgThree}\\
\verb|find_proof HamCheck|
\vspace*{1em}\\
\noindent
We made this example simple, so that
two sub-goals, \pgOne{} and \pgThree{}, are not hard to prove;
however,
they are still beyond the scope of commonly used tactics, such as \fastforce{}.

Generally, for a conjecture and a strategy of the form of \verb|RepeatN (strategy)|,
\lang{} applies \strategy{} to the conjecture
as many times as the number of proof obligations in the conjecture.
In this case, \lang{} applies \texttt{Ors [Hammer, \checkpickdef{}]} to \lemmTwo{} three times.

Note that we integrated \quickcheck{} and \nitpick{} into \lang{} as \textit{assertion tactics}.
Assertion tactics provide mechanisms for controlling proof search based on a condition:
such a tactic takes a proof state, tests an assertion on it, then behaves as \succeed{}
or \fail{} accordingly.
We have already seen one of them in the previous examples: \issolved{}.

\texttt{Ors [Hammer, \checkpickdef{}]} first applies \sledgehammer.
If \sledgehammer{} does not find a proof,
it tries to find counter-examples for the sub-goal using \quickcheck{}.
If \quickcheck{} finds no counter-examples,
\lang{} interprets \Defer{} as \verb|defer_tac 1|,
which postpones the current sub-goal to the end of the list of proof obligations.

In this example, \sledgehammer{} fails to discharge \pgTwo{}.
When \sledgehammer{} fails, \lang{} passes \two{} to \checkpickdef{},
which finds no counter-example to \two{} and sends \two{} to the end of the list;
then, \lang{} continues working on the sub-goal \texttt{3} with \sledgehammer{}.
The runtime stops its execution after applying
\verb|Ors [Hammers,| \checkpickdef{}\verb|]| three times,
generating the following proof script.
This script discharges \texttt{1} and \texttt{3},
but it leaves \two{} as the meaningful task for human engineers,
while assuring there is no obvious counter-examples for \two{}.
\begin{verbatim}
apply (simp add: state_safety ps_safe_def)
defer apply (simp add: state_safety ps_safe_def)
\end{verbatim}
\section{The default strategy: \tryhard{}.}\label{s:tryhard}

\lang{} comes with a default strategy, \tryhard{}.
Users can apply \tryhard{} as a completely automatic tool:
engineers need not provide their intuitions by writing strategies.
Unlike other user-defined strategies,
one can invoke this strategy by simply typing \tryhard{}
without \findproof{} inside a proof attempt.
The lack of input from human engineers makes \tryhard{} less specific to each conjecture;
however, we made \tryhard{} more powerful than existing proof automation tools for Isabelle
by specifying larger search spaces presented in Appendix A.

\begin{table}[t]
\caption{The number of automatically proved proof obligations from assignments.}\label{table:eval_hw}
\begin{center}
\begin{tabular}{ccccccccccc}
\hline\noalign{\smallskip}
assignments \cite{CSCourse}
& ass\_1  & ass\_2 & ass\_3 & ass\_4 & ass\_5 & ass\_6 & ass\_8 & ass\_9 & ass\_11 & sum\\
\noalign{\smallskip}
\verb|POs| & 19  & 22 & 52 & 82 & 64 & 26 & 52 & 61 & 26 & 404\\
\hline
\noalign{\smallskip}
\verb|TH| 30s     & 17      & 21   & 30   & 66   & 36     & 11    & 36    & 31     & 14 & 262\\
\verb|SH| 30s     & 14(13)  & 5    & 27   & 61   & 41(39) & 12(11)& 45(39)& 32(30) & 15 & 252(241)\\
\verb|TH\SH| 30s  & 4       & 16   &  8   & 10   & 6      & 2     & 1     & 6      &  1 & 54\\
\verb|TH| 300s    & 18      & 22   & 35   & 71   & 55     & 14    & 40    & 35     & 20 & 310\\
\verb|SH| 300s    & 14(13)  & 5    & 27   & 61   & 44(42) & 13(12)& 46(39)& 32(30) & 17 & 259(246)\\
\verb|TH\SH| 300s & 5       & 17   & 10   & 10   & 17     & 3     & 0     & 6      & 3  & 71\\
\hline
\end{tabular}
\end{center}
\end{table}

\begin{table}[t]
\caption{The number of automatically proved proof obligations from exercises.}\label{table:eval_ex}
\begin{center}
\begin{tabular}{ccccccccccccccccc}
\hline\noalign{\smallskip}
exercise \cite{CSCourse}
& e1  & e2 & e3 & e4 & e5a & e5b & e6 & e7a & e7b & e8a & e8b & e9 & e10 & e11 & e12 & sum\\
\noalign{\smallskip}
\verb|POs| & 15 & 7 & 42 & 23 & 13 & 83 & 4 & 3 & 9 & 10 & 26 & 31 & 15 & 10 & 30 & 321\\
\hline
\noalign{\smallskip}
\verb|TH| 30s     & 12 & 4 & 27     & 11 & 9  & 65 & 1 & 0 & 5 & 7 & 11 & 14 & 5    & 4 & 8  & 183\\
\verb|SH| 30s     & 8  & 3 & 26(25) & 15 & 11 & 74 & 2 & 0 & 6 & 7 & 9  & 17 & 5(4) & 6 & 10 & 199 (197)\\
\verb|TH\SH| 30s  & 4  & 2 & 5      & 0  & 0  & 1  & 0 & 0 & 1 & 1 & 4  & 3  & 1    & 0 & 1  & 23\\
\verb|TH| 300s    & 12 & 5 & 29     & 17 & 11 & 74 & 1 & 0 & 8 & 7 & 12 & 19 & 6    & 9 & 12 & 222\\
\verb|SH| 300s    & 8  & 3 & 27(26) & 15 & 11 & 74 & 3 & 0 & 6 & 7 & 12 & 17 & 6(5) & 6 & 10 & 205(203)\\
\verb|TH\SH| 300s & 4  & 2 & 5      & 2  & 0  & 1  & 0 & 0 & 2 & 1 & 2  & 3  & 1    & 3 & 3  & 29\\
\hline
\end{tabular}
\end{center}
\end{table}

\begin{table}[t]
\caption{The number of automatically proved proof goals from AFP entries and Isabelle's standard libraries.}
\label{table:eval_afp}
\begin{center}
\begin{tabular}{cccccccc}
\hline\noalign{\smallskip}
theory name & \verb|POs| & \verb|TH| & \verb|SH| & \verb|TH\SH| & \verb|TH| & \verb|SH| & \verb|TH\SH|\\
\hline\noalign{\smallskip} time out & - & 30s & 30s & 30s & 300s & 300s & 300s\\
\hline
\noalign{\smallskip}
\verb|DFS.thy| \cite{Depth-First-Search-AFP}                     & 51  & 24  & 28     &  6  & 34  & 29      & 7\\
\verb|Efficient_Sort.thy| \cite{Efficient-Mergesort-AFP}         & 75  & 27  & 28(26) &  8  & 33  & 31(28)  & 9\\
\verb|List_Index.thy| \cite{List-Index-AFP}                      & 105 & 48  & 72(70) &  12 & 67  & 75(72)  & 14\\
\verb|Skew_Heap.thy| \cite{Skew_Heap-AFP}                        & 16  & 8   & 6(5)   &  4  & 12  & 8(7)    & 5\\
\verb|Hash_Code.thy| \cite{Collections-AFP}                      & 16  & 7   & 4      &  4  & 11  & 4       & 7\\
\verb|CoCallGraph.thy| \cite{Call_Arity-AFP}                     & 141 & 88  & 78(71) &  29 & 104 & 79(73)  & 33\\
\verb|Coinductive_Language.thy| \cite{Coinductive_Languages-AFP} & 139 & 57  & 69(68) &  11 & 106 & 70(69)  & 43\\
\verb|Context_Free_Grammar.thy| \cite{Coinductive_Languages-AFP} & 29  & 26  & 2      &  26 & 29  & 2       & 27\\
\verb|LTL.thy| \cite{LTL-AFP}                                    & 97  & 56  & 61     &  15 & 78  & 65(62)  & 15\\
\verb|HOL/Library/Tree.thy|                                      & 124 & 93  & 70(68) &  32 & 101 & 73(70)  & 32\\
\verb|HOL/Library/Tree_Multiset.thy|                             & 8   & 8   & 1      &  7  & 8   & 1       & 7\\
     sum                                                         & 801 & 442 &419(404)& 154 & 583 & 437(417)& 199\\
\hline
\end{tabular}
\end{center}
\end{table}

We conducted a judgement-day style evaluation \citep{DBLP:conf/cade/BohmeN10} of \tryhard{}
against selected theory files from the Archive of Formal Proofs (AFP),
coursework assignments and exercises \citep{CSCourse}, and Isabelle's standard library.
Table 1, 2 and 3
\footnote{\texttt{TH} and \texttt{SH} stand for
the number of obligations discharged by \tryhard{} and \sledgehammer{}, respectively.
\texttt{TH\textbackslash SH} represents the number of goals 
to which \tryhard{} found proofs but \sledgehammer{} did not.
\texttt{POs} stands for the number of proof obligations in the theory file.
$x(y)$ for \texttt{SH} means \sledgehammer{} found proofs for $x$ proof obligations, out of which it managed to reconstruct proof scripts in Isabelle for $y$ goals.
We omit these parentheses when these numbers coincide.
Note that all proofs of \lang{} are checked by Isabelle/HOL.
Besides, \sledgehammer{} inside \lang{} avoids the \texttt{smt} proof method,
as this method is not allowed in the Archive of Formal Proofs.
}
show that given 300 seconds of time-out for each proof goal
\tryhard{} solves 1115 proof goals out of 1526,
while \sledgehammer{} found proofs for 901 of them 
using the same computational resources
and re-constructed proofs in Isabelle for 866 of them.
This is a 14 percentage point improvement of proof search
and a 16 percentage point increase for proof reconstruction.
Moreover, 299 goals (20\% of all goals) were solved only by \tryhard{} within 300 seconds.
They also show that a longer time-out improves the success ratio of \tryhard{}, which is desirable for utilising engineers' downtime.

\tryhard{} is particularly more powerful than \sledgehammer{} at discharging proof obligations that can be nicely handled by the following:
\begin{itemize}[noitemsep,nolistsep]
  \item mathematical induction or co-induction,
  \item type class mechanism,
  \item specific procedures implemented
        as specialised tactics (such as
        \transfer{} and \verb|normalization|), or
  \item general simplification rules (such as \verb|field_simps| and \verb|algebra_simps|) .
\end{itemize}

Furthermore,
careful observation of \lang{} indicates that
\lang{} can handle the so-called ``hidden-fact'' problem of relevance filter.
``Hidden facts'' are auxiliary lemmas
that are useful to discharge a given proof obligation
but hard to pick up for a relevance filter
because they are ``hidden'' in the definition of other facts.
With \lang{}, users can write strategies,
which first applies rewriting to a given conjecture to reveal these hidden facts to
the relevance filter.
For example, the following strategy ``massages'' the given proof obligation before invoking
the relevance filter of \sledgehammer{}:
\verb|Thens| \verb|[Auto,| \verb|Repeat(Hammer),| \verb|IsSolved]|.

For 3 theories out of 35,
\tryhard{} discharged fewer proof obligations,
even given 300 seconds of time-out.
This is due to the fact that
\lang{} uses a slightly restricted version of \sledgehammer{} internally
for the sake of the integration with other tools
and to avoid the \verb|smt| method, which is not allowed in the AFP.
In these files,
\sledgehammer{} can discharge many obligations and
other obligations are not particularly suitable for other sub-tools in \tryhard{}.
Of course, given high-performance machines,
users can run both \tryhard{} and \sledgehammer{} in parallel
to maximise the chance of proving conjectures.

\section{Monadic Interpretation of Strategy}\label{s:psm}
The implementation of the tracing mechanism described in Section \ref{s:example}
is non-trivial:
\lang{}'s tracing mechanism has to support arbitrary strategies conforming to its syntax.
What is worse,
the runtime behaviour of backtracking search is not completely predictable statically
since \lang{} generates tactics at runtime,
using information that is not available statically.
Moreover, the behaviour of each tactic varies
depending on the proof context and proof obligation at hand.

Implementation based on references or pointers is likely to cause code clutter,
whereas explicit construction of search tree \cite{Nagashima_16} consumes too much memory space when traversing a large search space.
Furthermore, both of these approaches deviate from the standard execution model of Isabelle
explained in Section \ref{background},
which makes the proof search and the efficient proof script generation less reliable.
In this section, we introduce our monadic interpreter for \lang{},
which yields a modular design and concise implementation of
\lang{}'s runtime system.

\paragraph{Monads in Standard ML.}
\begin{program}[t]
\begin{verbatim}

signature MONAD0PLUS =
sig  type 'a m0p;
     val return :  'a -> 'a m0p;
     val bind   :  'a m0p -> ('a -> 'b m0p) -> 'b m0p;
     val mzero  :  'a m0p;
     val ++     : ('a m0p * 'a  m0p) -> 'a m0p;
end;
structure Nondet : MONAD0PLUS =
struct type 'a m0p     = 'a Seq.seq;
       val return      = Seq.single;
       fun bind xs f   = Seq.flat (Seq.map f xs);
       val mzero       = Seq.empty;
       fun (xs ++ ys)  = Seq.append xs ys;
end;
\end{verbatim}
\caption{Monad with zero and plus, and lazy sequence as its instance.}
\label{p:m0p}
\end{program}
\label{program:mis}

A monad with zero and plus is a constructor class\footnote{
Constructor classes are a class mechanism on type constructors.}
with four basic methods.
As Isabelle's implementation language, Standard ML, does not natively support
constructor classes,
we emulated them using its module system \cite{Nagashima_OConnor_16}.
Program \ref{p:m0p} shows how we represent the type constructor, \verb|seq|,
as an instance of monad with zero and plus.

The body of \verb|bind| for lazy sequences says that
it applies \verb|f| to all the elements of \verb|xs| and
concatenates all the results into one sequence.
Attentive readers might notice that this is equivalent to
the behaviour of \THEN{} depicted in Fig. \ref{fig:internal}
and that of \Thens{} shown in Fig. \ref{fig:dynamic_induct}.
In fact, we can define all of
\THEN{}, \succeed{}, \fail{}, and \APPEND{},
using \verb|bind|, \verb|return|, \verb|mzero|, and \verb|++|, respectively.

\paragraph{Monadic Interpretation of Strategies.}
Based on this observation, we formalised
\lang{}'s search procedure as a monadic interpretation of strategies, as shown in Program \ref{p:interp},
where the type \verb|core_strategy| stands for the internal representation of strategies.
Note that \Alt{} and \Or{} are binary versions of \Alts{} and \Ors{}, respectively;
\lang{} desugars \Alts{} and \Ors{} into nested \Alt{}s and \Or{}s.
We could have defined \Or{} as a syntactic sugar
using \Alt, \mzero, \Fail, and \Skip, as explained
by Martin \etal{} \cite{DBLP:journals/fac/MartinGW96};
however, we prefer the less monadic formalisation in Program \ref{p:interp} for better time complexity.
\begin{program}[t]
\begin{verbatim}

interp :: core_strategy -> 'a -> 'a m0p
interp (Atom atom_str) n     = eval atom_str n
interp  Skip  n              = return n
interp  Fail  n              = mzero
interp (str1 Then str2) n    = bind (interp str1 n) (interp str2)
interp (str1 Alt  str2) n    = interp str1 n ++ interp str2 n
interp (str1 Or   str2) n    = let val result1 = interp str1 n
  in if (result1 != mzero) then result1 else interp str2 n end
interp (Rep str) n           = interp ((str THEN (Rep str)) Or Skip) n
interp (Comb (comb, strs)) n = eval_comb (comb, map interp strs) n
\end{verbatim}
  \caption{The monadic interpretation of strategies.}
  \label{p:interp}
\end{program}

\eval{} handles all the atomic strategies, which correspond to
\verb|default|, \verb|dynamic|, and \verb|special| in the surface language.
For the \verb|dynamic| strategies,
\eval{} expands them into dynamically generated tactics
making use of contextual information from the current proof state.
\lang{} combines these generated tactics either with \APPEND{} or \ORELSE{},
depending on the nature of each tactic.
\verb|eval_comb| handles non-monadic strategy combinators, such as \verb|Cut|.
We defined the body of \eval{} and \verb|eval_comb| for each atomic strategy and strategy combinator separately using pattern matching.
As is obvious in Program \ref{p:interp},
\verb|interp| separates the complexity of compound strategies
from that of runtime tactic generation.

\paragraph{Adding Tracing Modularly for Proof Script Generation.}
We defined \verb|interp| at the constructor class level,
abstracting it from the concrete type of proof state and
even from the concrete type constructor.
When instantiated with lazy sequence,
\verb|interp| tries to return the first element of the sequence,
working as depth-first search.
This abstraction provides a clear view of how compound strategies guide proof search
while producing tactics at runtime;
however, without tracing proof attempts,
\lang{} has to traverse large search spaces every time it checks proofs.

\begin{program}[t]
\begin{verbatim}

functor writer_trans (structure Log:MONOID; structure Base:MONAD0PLUS) =
struct type 'a m0p  = (Log.monoid * 'a) Base.m0p;
       fun return (m:'a) = Base.return (Log.mempty, m) : 'a m0p;
       fun bind (m:'a m0p) (func: 'a -> 'b m0p) : 'b m0p =
          Base.bind    m          (fn (log1, res1) =>
          Base.bind   (func res1) (fn (log2, res2) =>
          Base.return (Log.mappend log1 log2, res2)));
       val mzero = Base.mzero;
       val (xs ++ ys) = Base.++ (xs, ys);
end : MONAD0PLUS;
\end{verbatim}
  \caption{The writer monad transformer as a ML functor.}
  \label{p:writer_trans}
\end{program}
We added the tracing mechanism to \verb|interp|,
combining the non-deterministic monad, \verb|Nondet|, with the writer monad.
To combine multiple monads,
we emulate monad transformers using ML functors:
Program \ref{p:writer_trans} shows
our ML functor, \verb|writer_trans|, which takes a module of \verb|MONAD0PLUS|,
adds the logging mechanism to it, and returns a module equipped with both
the capability of the base monad and the logging mechanism of the writer monad.
We pass \verb|Nondet| to \verb|writer_trans| as the base monad
to combine the logging mechanism and
the backtracking search based on non-deterministic choice.
Observe Program \ref{p:m0p}, \ref{p:interp} and \ref{p:writer_trans} to see
how \verb|Alt| and \verb|Or| truncate failed proof attempts while searching for a proof.
The returned module is based on a new type constructor,
but it is still a member of \verb|MONAD0PLUS|;
therefore, we can re-use \verb|interp| without changing it.

\paragraph{History-Sensitive Tactics using the State Monad Transformer.}
The flexible runtime interpretation might lead \lang{} into a non-terminating loop,
such as \texttt{REPEAT succeed}.
To handle such loops, \lang{} traverses a search space
using iterative deepening depth first search (IDDFS).
However, passing around information about depth
as an argument of \verb|interp| as following\footnote{
\texttt{level} stands for the remaining depth
\texttt{interp} can proceed for the current iteration.}
quickly impairs its simplicity.
\begin{verbatim}
interp (t1 CSeq t2) level n = if level < 1 then return n else ...
interp (t1 COr  t2) level n = ...
\end{verbatim}
We implemented IDDFS without code clutter,
introducing the idea of a \textit{history-sensitive tactic}:
a tactic that takes the log of proof attempts into account.
Since the writer monad does not allow us to access the log during the search time,
we replaced the writer monad transformer with the state monad transformer,
with which the runtime keeps the log of proof attempt as the ``state'' of proof search and access it during search.
By measuring the length of ``state'',
\verb|interp| computes the current depth of proof search at runtime.

The modular design and abstraction discussed above made this replacement possible
with little change to the definition of \verb|interp|:
we only need to change the clause for \verb|Atom|,
providing a wrapper function, \verb|iddfc|, for \eval{},
while other clauses remain intact.
\begin{verbatim}
inter (CAtom atom_str) n = iddfc limit eval atom_str n
\end{verbatim}
\verb|iddfc limit| first reads the length of ``state'',
which represents the number of edges to the node from the top of the implicit proof search tree.
Then, it behaves as \fail{} if the length exceeds \verb|limit|;
if not, it executes \verb|eval| \verb|atom_str| \verb|n|.\footnote{
In this sense, we implemented IDDFS as a tactic combinator.}

\section{Related Work}
\ACL{} \cite{Kaufmann:2000:CRA:555902} is a functional programming language
and mostly automated first-order theorem prover,
while \lang{} is embedded in Isabelle/HOL to support higher-order logic.
\ACL{} is known for the so-called waterfall model,
which is essentially repeated application of various heuristics.
Its users can guide proof search by supplying arguments called ``hints'',
but the underlining operational procedure of the waterfall model itself is fixed.
\ACL{} does not produce efficient proof scripts after running the waterfall algorithm.

\PVS{} \cite{Owre:1992:PPV:648230.752639} provides a collection of commands called ``strategies''.
Despite the similarity of the name to \lang{},
strategies in PVS correspond to tactics in Isabelle.
The highest-level strategy in \PVS{},
\grind{}, can produce re-runnable proof scripts containing successful proof steps only.
However, scripts returned by \grind{} describe steps of much lower level
than human engineers would write manually,
while \lang{}'s returned scripts are based on tactics engineers use.
Furthermore, \grind{} is known to be useful to complete a proof
that does \textit{not} require induction,
while
\tryhard{} is good at finding proofs
involving mathematical induction.

\SEPIA{} \cite{DBLP:conf/cade/GransdenWR15} is an automated proof generation tool in Coq.
Taking existing Coq theories as input,
\SEPIA{} first produces proof traces,
from which it infers an extended finite state machine. 
Given a conjecture, \SEPIA{} uses this model to search for its proof.
\SEPIA{}'s search algorithm is based on the breadth-first search (BFS)
to return the shortest proof.
\lang{} can also adopt BFS,
as BFS is a special case of IDDFS.
However,
our experience tells that the search tree tends to be very wide
and some tactics, such as \induct{}, need to be followed by other tactics to complete proofs.
Therefore, we chose IDDFS for \lang{}.
Both \SEPIA{} and \lang{} off-load the construction of proof scripts to search
and try to reconstruct efficient proof scripts.
Compared to \SEPIA, \lang{} allows users to specify their own search strategies
to utilize the engineer's intuition,
which enables \lang{} to return incomplete proof scripts,
as discussed in Section \ref{incomplete_proof}.

Martin \etal{} first discussed a monadic interpretation of tactics
for their language, \textit{Angel},
in an unpublished draft \cite{Martin02amonadic}.
We independently developed \verb|interp| with the features discussed above,
lifting the framework from the tactic level to the strategy level
to traverse larger search spaces.
The two interpreters for different ITPs turned out to be similar to each other,
suggesting
our approach is not specific to Isabelle but can be used for other ITPs.

Similar to \textit{Ltac} \citep{DBLP:conf/lpar/Delahaye00} in Coq,
\Eisbach{} \citep{DBLP:conf/itp/MatichukWM14} is a framework to write
\textit{proof method}s in Isabelle. 
Proof methods are the \Isar{} syntactic layer of tactics.
Eisbach does not generate methods dynamically, trace proof attempts, nor support parallelism natively.
Eisbach is good when engineers already know how to prove their conjecture,
while \texttt{try\_hard} is good when they want to find out how to prove it.

\IsaPlanner{} \citep{DBLP:conf/cade/DixonF03} offers a framework
for encoding and applying common patterns of reasoning in Isabelle,
following the style of proof planning \cite{DBLP:conf/cade/Bundy88}.
\IsaPlanner{} addresses the performance issue by memoization technique,
on the other hand \texttt{try\_hard} strips off backtracked steps
while searching for a proof,
which Isabelle can check later without \texttt{try\_hard}.
While \IsaPlanner{} works on its own data structure \textit{reasoning state},
\tryhard{} managed to minimize the deviation from Isabelle's standard execution model
using constructor classes.

\section{Conclusions}\label{s:conc}
Proof automation in higher-order logic is a hard problem:
every existing tool has its own limitation.
\lang{} attacks this problem, allowing
us to exploit both the engineer's intuition and various automatic tools.
The simplicity of the design is our intentional choice:
we reduced the process of interactive proof development
to the well-known dynamic tree search problem
and added new features (efficient-proof script generation and IDDFS)
by safely abstracting the original execution model and
employing commonly used techniques (monad transformers).

We claim that
our approach enjoys significant advantages.
Despite the simplicity of the design,
our evaluations indicate that \lang{} reduces the labour cost of ITP significantly.
The conservative extension to the original model lowers the learning barrier of \lang{}
and makes our proof-script generation reliable by minimising the deviation.
The meta-tool approach makes the generated proof-script independent of \lang{},
separating the maintenance of proof scripts from that of \lang{};
furthermore, by providing a common framework for various tools
we supplement one tool's weakness (e.g. induction for \sledgehammer{})
with other tools' strength (e.g. the \induct{} tactic),
while enhancing their capabilities with runtime tactic generation.
The parallel combinators transforms
the conventionally labour-intensive interactive theorem proving
to embarrassingly parallel problems.
The abstraction to the constructor class and
reduction to the tree search problem make our ideas transferable:
other ITPs, such as Lean and Coq, handle inter-tactic backtracking,
which is best represented in terms of \verb|MONAD0PLUS|.

\subsubsection*{Acknowledgements}
We thank Jasmin C. Blanchette for his extensive comments that improved
the evaluation of \tryhard{}.
Pang Luo helped us for the evaluation.
Leonardo de Moura, Daniel Matichuk, Kai Engelhardt, and Gerwin Klein provided valuable comments on an early draft of this paper.

\newpage
  \bibliographystyle{splncs03}
  \bibliography{refs}

\pagebreak

\newpage
\appendix
\section{Appendix: the Default Strategy, \tryhard{}}\label{s:tryhard}
The following is the definition of \tryhard{}.
It starts with simple sub-strategies and gradually proceeds to more involved sub-strategies.
Note that \tryhard{} is just one default strategy:
we provided \lang{} as a language,
so that users can encode their intuitions as strategies.
\begin{verbatim}
strategy Auto_Solve   = Thens [Auto, IsSolved]
strategy Blast_Solve  = Thens [Blast, IsSolved]
strategy FF_Solve     = Thens [Fastforce, IsSolved]
strategy Auto_Solve1  = Thens [Subgoal, Auto, IsSolved]
strategy Auto_Hammer  = Thens [Subgoal, Auto, RepeatN(Hammer), 
                               IsSolved]
strategy Solve_One    = Ors [Fastforce, Auto_Solve1, Hammer]
strategy Solve_Many   = Thens [Repeat (Solve_One), IsSolved]
strategy DInduct      = Dynamic (Induct)
strategy DInductTac   = Dynamic (InductTac)
strategy DCoinduction = Dynamic (Coinduction)
strategy DCases       = Dynamic (Cases)
strategy DCaseTac     = Dynamic (CaseTac)
strategy DAuto        = Dynamic (Auto)
strategy Basic =
  Ors [Auto_Solve,
       Blast_Solve,
       FF_Solve,
       Thens [IntroClasses, Auto_Solve],
       Thens [Transfer, Auto_Solve],
       Thens [Normalization, IsSolved],
       Thens [DInduct, Auto_Solve],
       Thens [Hammer, IsSolved],
       Thens [DCases, Auto_Solve],
       Thens [DCoinduction, Auto_Solve],
       (*Occasionally, auto reveals hidden facts.*)
       Thens [Auto, RepeatN(Hammer), IsSolved],
       Thens [DAuto, IsSolved]]
strategy Advanced =
  Ors [Solve_Many,
       Thens [DCases, DCases, Auto_Solve],
       Thens [DCases, Solve_Many],
       Thens [IntroClasses,
              Repeat (Ors[Fastforce, 
                          Thens[Transfer, Fastforce], Solve_Many]),
              IsSolved],
       Thens [Transfer, Solve_Many],
       Thens [DInduct, Solve_Many],
       Thens [DCoinduction, Solve_Many]]
strategy Try_Hard =
  Ors [Thens [Subgoal, Basic],
       Thens [DInductTac, Auto_Solve],
       Thens [DCaseTac, Auto_Solve],
       Thens [Subgoal, Advanced],
       Thens [DCaseTac, Solve_Many],
       Thens [DInductTac, Solve_Many]]
\end{verbatim}

\section{Appendix: Details of the Evaluation}
All evaluations were conducted on a Linux machine with
Intel (R) Core (TM) i7-600 @ 3.40GHz and 32 GB memory.
For both tools, we set the time-out of proof search to 30 and 300 seconds for each proof obligation.

Prior to the evaluation, the relevance filter of \sledgehammer{} was
trained on 27,041 facts and 18,695 non-trivial Isar proofs
from the background libraries imported by theories under evaluation for both tools.
Furthermore, we forbid \sledgehammer{} inside \lang{} from using
the \verb|smt| method for proof reconstruction,
since the AFP does not permit this method.

Note that \tryhard{} does not use parallel strategy combinators which exploit parallelism.
The evaluation tool does not allow \tryhard{} to use multiple threads either.
Therefore, given the same time-out, \tryhard{} and \sledgehammer{} enjoy 
the same amount of computational resources,
assuring the fairness of the evaluation results.

We provide the evaluation results 
in the following website
for the purpose of reviewing:

\begin{itemize}
\item \url{http://ts.data61.csiro.au/Private/downloads/cade26_results/}
\end{itemize}

\section{Appendix: Note on Examples.}
The examples of efficient-proof-script generation in Section \ref{s:example} have been demonstrated on machines at Data61.
Depending on the hardware conditions and the prior training of \sledgehammer{}'s relevance filter,
one may obtain different results.
One can configure the number of threads for proof search
by inserting the following ML code snippet inside the ongoing Isabelle proof script:
\begin{verbatim}
ML{* Multithreading.max_threads_update 56 *}
ML{* Goal.parallel_proofs := 0 *}
\end{verbatim}

\end{document}